\documentclass[12pt]{article}

\usepackage{graphicx}
\usepackage[nomarkers,figuresonly]{endfloat} 
\usepackage{amssymb}
\usepackage{amsmath}
\usepackage{authblk}
\usepackage{color} 
\usepackage[utf8x]{inputenc} 

\usepackage[sort&compress]{natbib}
\newcommand{\Deg}{\ensuremath{^\circ}} 

\title{Quantitative analysis of magnetic spin and orbital moments from an oxidized iron (1~1~0) surface using electron magnetic circular dichroism} 

\author[1]{Thomas~Therslef\mbox{}f\thanks{Corresponding author.  Email: thomas.thersleff@angstrom.uu.se}}
\author[2]{Jan~Rusz}
\author[1,3]{Stefano~Rubino}
\author[2]{Bj{\"o}rgvin~Hj{\"o}rvarsson}
\author[4]{Yasuo~Ito}
\author[5]{Nestor~Zaluzec}
\author[1]{Klaus~Leifer\thanks{Corresponding author.  Email: klaus.leifer@angstrom.uu.se}}

\affil[1]{\small Department of Engineering Sciences, Uppsala University, Uppsala, Sweden}
\affil[2]{Department of Physics and Astronomy, Uppsala University, Uppsala, Sweden}
\affil[3]{Department of Physics, University of Oslo, Oslo, Norway}
\affil[4]{Department of Physics, Northern Illinois University, DeKalb, IL, USA}
\affil[5]{Electron Microscopy Center, Argonne National Laboratory, Argonne, IL, USA}

\begin{document}
\maketitle

\begin{abstract}
Understanding the ramifications of reduced crystalline symmetry on magnetic behavior is a critical step in improving our understanding of nanoscale and interfacial magnetism.  However, investigations of such effects are often controversial largely due to the challenges inherent in directly correlating nanoscale stoichiometry and structure to magnetic behavior.  Here, we describe how to use Transmission Electron Microscope (TEM) to obtain Electron Magnetic Circular Dichroism (EMCD) signals as a function of scattering angle to locally probe the magnetic behavior of thin oxide layers grown on an Fe (1~1~0) surface.  Experiments and simulations both reveal a strong dependence of the magnetic orbital to spin ratio on its scattering vector in reciprocal space.  We exploit this variation to extract the magnetic properties of the oxide cladding layer, showing that it locally may exhibit an enhanced orbital to spin moment ratio.  This finding is supported here by both spatially and angularly resolved EMCD measurements, opening up the way for compelling investigations into how magnetic properties are affected by nanoscale features.
\end{abstract}


\label{Introduction}

Systems of restricted size and dimensionality represent a frontier for research on magnetic materials.  By reducing the spatial dimensions of crystalline magnets, interfacial magnetic properties become more prominent in the overall magnetic behavior of the system \cite{gambardella_ferromagnetism_2002}.  Controlling interfaces through the fabrication of magnetic heterostructures enables researchers to produce materials that exhibit entirely new properties from the bulk constituents alone.  For these reasons, considerable research efforts are currently dedicated to improving the understanding of nanoscale magnetic behavior; however, many of the techniques capable of quantifying magnetic moments - such as \mbox{X-ray} Magnetic Circular Dichroism (XMCD) - lack the spatial resolution necessary to directly correlate the magnetic behavior of the material to the nanoscale features from which they arise.  

An example of a field where such understanding is necessary is research on magnetic transition metal oxides, particularly the iron oxides.  Iron oxides form in a variety of phases exhibiting a wide range of magnetic behavior.  Of particular interest is the phase Fe$_3$O$_4$, commonly known as magnetite.  Magnetite has potential to play a role in the development of nanoscale magnetic applications due to its anticipated half-metallic behavior \cite{katsnelson_half-metallic_2008} leading to a near 100\% spin polarization \cite{bibes_oxide_2007,wada_efficient_2010}, high chemical stability in ambient conditions, and ability to stabilize very thin films against the onset of superparamagnetism \cite{monti_magnetism_2012}.  It crystallizes in an inverse spinel cubic structure with Fe assuming both +2 and +3 oxidation states in a ferrimagnetic arrangement.  Fe$^{3+}$ is evenly distributed among the tetragonal and octahedral sites and is aligned antiparallel while Fe$^{2+}$ is found exclusively on the octahedral site, resulting in a net magnetic moment of nearly 4~$\mu_{\rm B}$. It is generally considered that the orbital component of this net magnetic moment in bulk magnetite is nearly completely compensated by this symmetry, resulting in a very small net contribution to the total magnetization \cite{duffy_spin_2010, goering_vanishing_2006, goering_absorption_2007}.  However, it has been suggested that local structural and chemical variations may break this compensation, increasing its detectable magnitude \cite{pellegrin_characterization_1999,huang_spin_2004, li_spin_2007, goering_large_2011, skoropata_magnetism_2014}.  Since these effects are necessarily nanoscale in origin while the measurement techniques employed probe macroscale materials, it has proven exceedingly difficult to explore these effects in greater detail.

One technique capable of contributing to this effort is Electron Magnetic Circular Dichroism (EMCD) in the Tranmission Electron Microscope (TEM).  First proposed in 2003 \cite{hebert_proposal_2003} and experimentally demonstrated in 2006 \cite{schattschneider_detection_2006}, EMCD is an exceptionally useful technique for quantifying the magnetic properties of materials on the nanometer scale \cite{schattschneider_detection_2008,lidbaum_reciprocal_2010,schattschneider_energy_2008} and has been employed in the analysis of magnetic domain walls at the nanoscale \cite{che_characterization_2011}, Magnetospirillum magnetotacticum \cite{stoger-pollach_emcd_2011}, LaSr-2~$\times$~4 manganese oxide nanowires \cite{carretero-genevrier_chemical_2012}, CrO$_2$ thin films \cite{loukya_electron_2012}, Fe$_3$O$_4$ nanoparticles \cite{salafranca_surfactant_2012}, and FeCo alloys \cite{warot-fonrose_magnetic_2010}.  The technique uses the TEM specimen as a beam splitter in the electron microscope where, in the correct scattering geometry, the fast electrons interact with the sample in a similar way to circularly polarized \mbox{x-rays}, enabling many of the measurements possible in XMCD to be executed in the TEM.  Whereas an XMCD signal is obtained from two spectra acquired using right and left-handed circularly polarized beams, in EMCD, two Electron Energy Loss Spectra (EELS) are acquired at two conjugated scattering vectors. The EMCD signal itself is the difference between these two spectra.

While the high spatial resolution of the EMCD technique clearly distinguishes it from XMCD, there are a number of additional differences that are just beginning to be explored and understood.  First, since the EMCD signal depends on the electron channeling conditions of the sample in the TEM, it is possible to use this technique to obtain site-specific magnetic information \cite{calmels_atomic_2011, wang_quantitative_2013}.  Second, it has been demonstrated that the magnitude of the EMCD signal (measured as its signal to noise ratio) varies as a function of scattering angle (or q-vector) \cite{rubino_energy-loss_2008, hebert_magnetic_2008, rubino_simulation_2010}.  This dependence - known as an ``EMCD strength map'' - can be experimentally recorded in the TEM either by angularly selecting the inelastically scattered electrons or by energy filtering a series of electron diffraction patterns.  When sum rules are applied to the resultant signals, it becomes possible to quantitatively extract the orbital to spin magnetic moment ratio $m_L / m_S$ as a function of q-space \cite{loukya_electron_2012, rusz_sum_2007, calmels_experimental_2007, lidbaum_quantitative_2009, rusz_local_2011}.  Significantly, while the magnitude of the EMCD signal changes as a function of crystalline symmetry of the magnetic material, the value of $m_L / m_S$ remains invariant for homogeneous systems \cite{lidbaum_quantitative_2009,rusz_influence_2011}

In this communication, we exploit both of these dynamical diffraction properties to extract $m_L / m_S$ for two magnetic materials having the same magnetic species but different crystal structures that overlap in the direction of the electron beam.  Such situations readily arise for metallic TEM lamellae prepared in cross section that have been exposed to atmosphere between sample preparation and transfer to the microscope, forming a thin metal oxide layer on the exposed surfaces.  If both the film as well as its oxide have a net magnetic moment, two independent, dissimilar EMCD strength maps will be generated.  To a good approximation, the total EMCD signal detected at any given q-vector will thus be a linear combination of these EMCD signal maps.  Since both EMCD strength maps vary differently in reciprocal space, when sum rules are applied to the entire EMCD signal map, the measured value for $m_L / m_S$ will vary as a function of detector position, which can be experimentally determined with high precision.

We exploit this effect to probe a TEM lamella of bcc iron having an exposed (1~1~0) surface, upon which a thin layer of cubic iron oxide has grown in the \{1~1~1\} orientation.  Based on structural and spectroscopy investigations, we argue that this layer is best described as Fe$_{3-\delta}$O$_4$ on both exposed surfaces, where $\delta$ varies between 0 (yielding mixed valence magnetite, Fe$_3$O$_4$) and 0.33 (yielding monovalent maghemite, $\gamma$--Fe$_2$O$_3$) \cite{pellegrin_characterization_1999}. By selectively probing different regions in reciprocal space, we show that it is possible to extract magnetic information pertaining to both the underlying iron film as well as its thin oxide surface layer.  Our experimental data combined with simulations on the system bring us to the conclusion that these thin Fe$_{3-\delta}$O$_4$ layers may locally exhibit a large, uncompensated orbital magnetic moment.

\section*{Results}
\label{Results}

\subsection*{Analysis of iron surface oxidation}

A film of bcc Fe with a thickness of 50~nm was epitaxially grown on a single crystalline MgO substrate as detailed in the methods section. An initial quality assessment of the film using \mbox{x-ray} diffraction techniques verified a close to single crystal (0~0~1) growth with a Full-Width at Half-Max (FWHM) of the rocking curve equal to 0.5\Deg.  A thin cross-sectional lamella was subsequently prepared for the TEM using the FIB in-situ lift-out method explained in the methods section.  An overview and structural assessment of the TEM-prepared sample is provided in the supplementary section.

\begin{figure}
	\includegraphics[width=\textwidth]{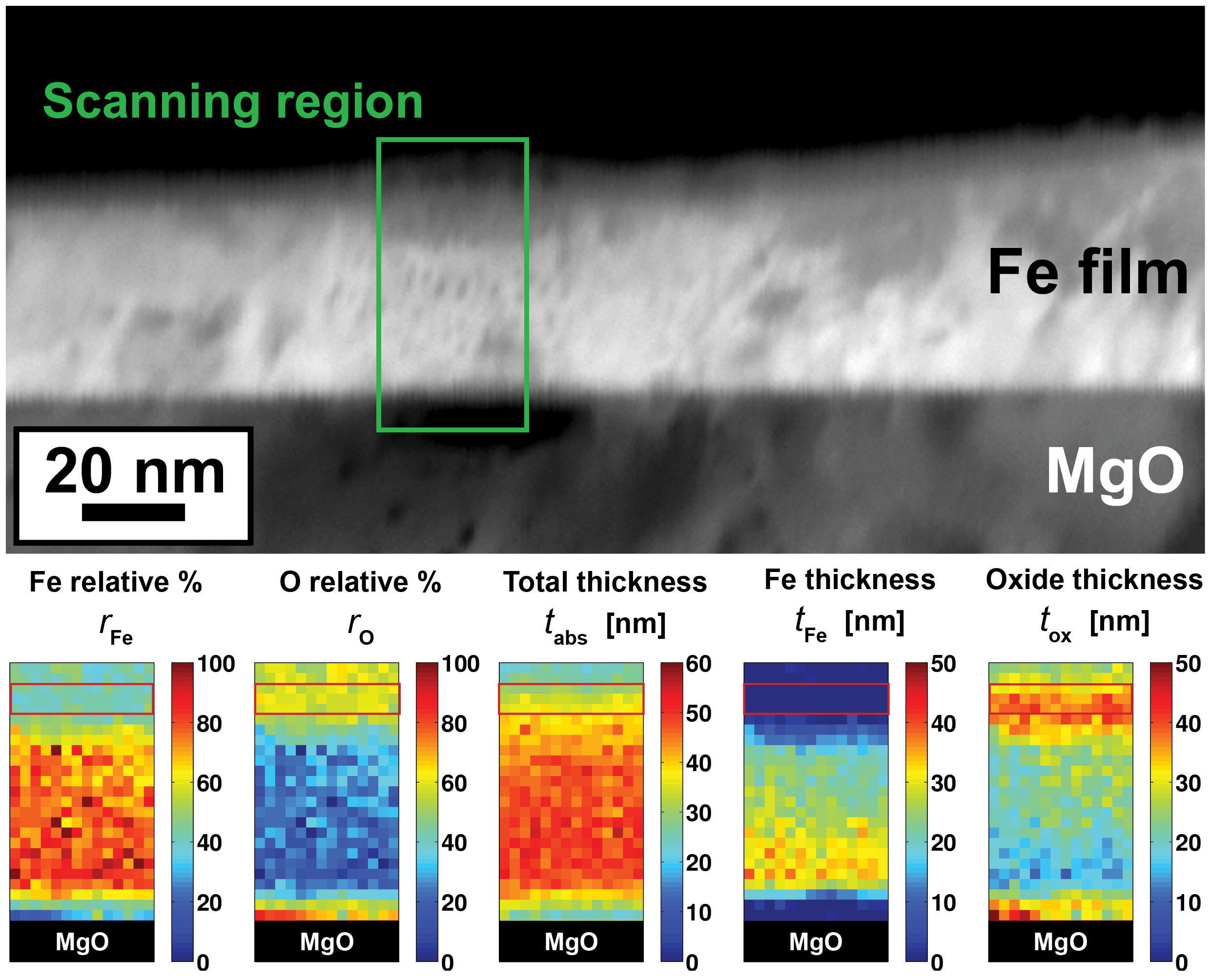} 
	\caption[Survey of the region of interest]{At top, an HAADF survey image denoting the region measured for EMCD is shown. The sub region denoted ``surface'' contains the largest volume percentage of oxide in the film and its position is marked as a red box on the maps.  Relative composition maps for Fe and O extracted from this region are shown in the lower left.  Absolute thickness maps for metallic Fe and Fe$_{3-\delta}$O$_4$ are shown in the lower right.  Note that the values for the absolute thickness in the MgO substrate region are blacked out since they cannot be accurately calculated with this method.  The pixel size of the maps is approximately 1.5 nm.}
	\label{fig:Survey}
\end{figure}

In the thinnest area of the lamella, a region of interest was sought out for the EMCD measurements.  A survey image of this region acquired with the microscope in Scanning TEM (STEM) mode using the High Angle Annular Dark Field (HAADF) detector is shown in f\mbox{}igure \ref{fig:Survey}.  The region used for this investigation is depicted in green.  The elemental composition of the iron cross-section in this region was investigated with on-axis EELS measurements as described in the methods section.  In the lower panel of figure \ref{fig:Survey}, real-space maps revealing the relative percentage of both Fe ($r_{\rm{Fe}}$) and O ($r_{\rm{O}}$) are presented along with the calculated total absolute thickness ($t_{\rm{abs}}$) of the film for any given pixel position.  The thickness of both the oxide cladding layers ($t_{\rm{ox}}$) as well as the underlying metallic layer ($t_{\rm{Fe}}$) was calculated under the assumption of the oxide being structurally similar to Fe$_3$O$_4$.  See the methods section for details about the thickness calculation.  

The on-axis EELS maps reveal that the entire Fe film within the scanned region exhibits a detectable oxygen signal.  In the middle of the Fe film, close to the substrate but slightly offset from the interface, oxygen comprises 10 -- 20\% of the total atomic concentration.  This rises to nearly 60\% in the upper region of the film, close to the vacuum. The absolute thickness of the film decreases as a function of distance from the MgO substrate, which is consistent with a wedge shape, as expected from the sample preparation procedure.  We therefore conclude that the oxide layer has encapsulated the original iron film and, towards the surface, has consumed nearly all of the metallic iron.  

Each pixel position in figure \ref{fig:Survey} contains both an individual EELS spectrum as well as spatial information with a resolution of approximately 1.5~nm. We note that spreading of the beam with convergence angle $\alpha = 1.6$~mrad during the propagation through a 50~nm sample thick is only approximately $50~\rm{nm} \cdot \alpha = 50~\rm{nm} \cdot 0.0016 < 0.1~\rm{nm}.$ This property allows for the EELS data to be spatially segregated into regions corresponding to significant local features.  Since the EMCD data were acquired from the same scanning region as the data in figure \ref{fig:Survey}, these core-loss EELS data can be used to determine the amount of oxide that contributes to the corresponding EMCD signal.  We thus identify two core regions of interest for the EMCD analysis.  The first is a summation over all spectra in each spectral image.  The second is a summation over the portion of the film with the highest oxide content, denoted the ``surface'' region.  The area of summation is presented in figure \ref{fig:Survey}.  

\begin{figure}
	\includegraphics[width=\linewidth]{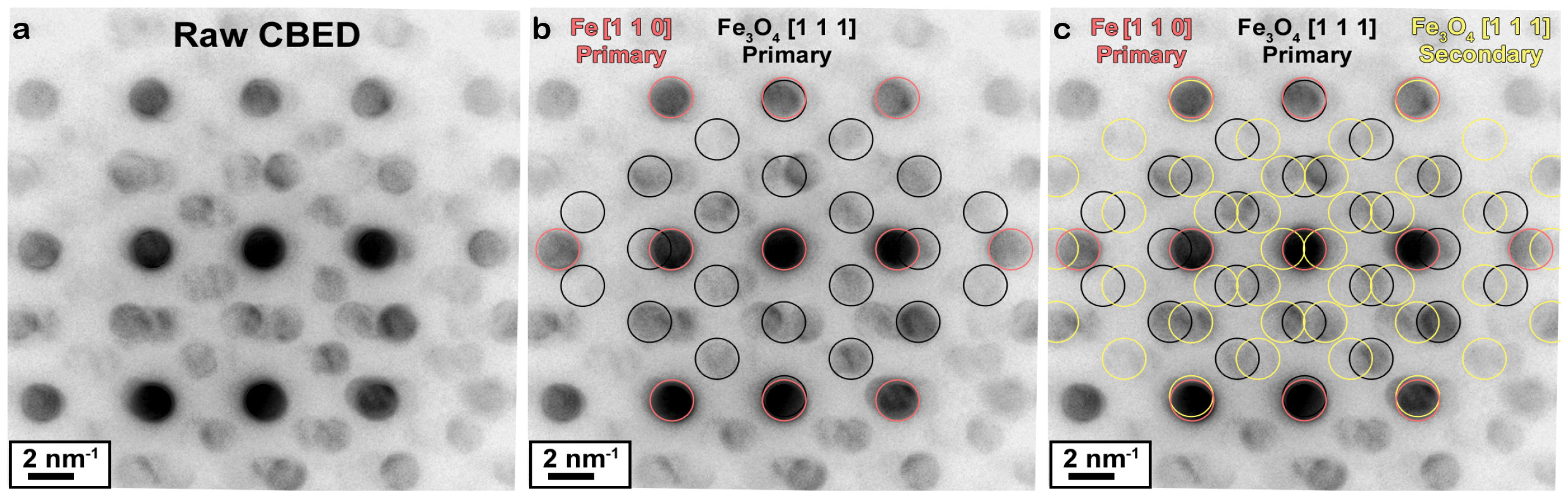} 
	\caption[CBED pattern from the region of interest]{CBED pattern from the scanned region (green box in figure \ref{fig:Survey}). Greyscale contrast is inverted to aid visualization in print.  In (a), the raw pattern is shown.  Low order indices for Fe (red) and Fe$_3$O$_4$ (black) are provided in (b) and secondary reflections arising from the interference mechanism are colored yellow in (c).}
	\label{fig:CBED}
\end{figure}

The structure of the oxide-encapsulated Fe layer was investigated with both Convergent Beam Electron Diffraction (CBED) and High Resolution TEM (HRTEM) techniques.  The CBED pattern is shown in figure \ref{fig:CBED}.  This pattern was extracted from the middle of the boxed region shown in figure \ref{fig:Survey} and thus reveals the structure of same area used for the EMCD measurements.  The most intense reflections can be indexed as Fe~[1~1~0], and this is done in figure \ref{fig:CBED}b.  The additional reflections appear to come in sets of three overlapping discs.  The midpoint of these sets of reflections matches well with the indices expected from stoichiometric Fe$_{3}$O$_4$ $[\overline{1}~\overline{1}~1]$, and these labels are used to index the structure in figure \ref{fig:CBED}b.  The satellite reflections can be understood as a consequence of dynamical diffraction effects resulting from the propagation of an electron wave through multiple crystalline lattices.  This produces an interference lattice that manifests itself as superposition of an additional spatial frequency over all of the primary Bragg reflections.  The frequency of this interference lattice $\Delta \mathbf{g}$ is 0.98~nm$^{-1}$ corresponding to $\mathbf{g}_{110}(\rm{Fe})-\mathbf{g}_{4\overline{2}2}(\rm{Fe}_3\rm{O}_4)$, and its lower orders are indexed in figure \ref{fig:CBED}c. 

\begin{figure}[p]
	\includegraphics[width=\textwidth]{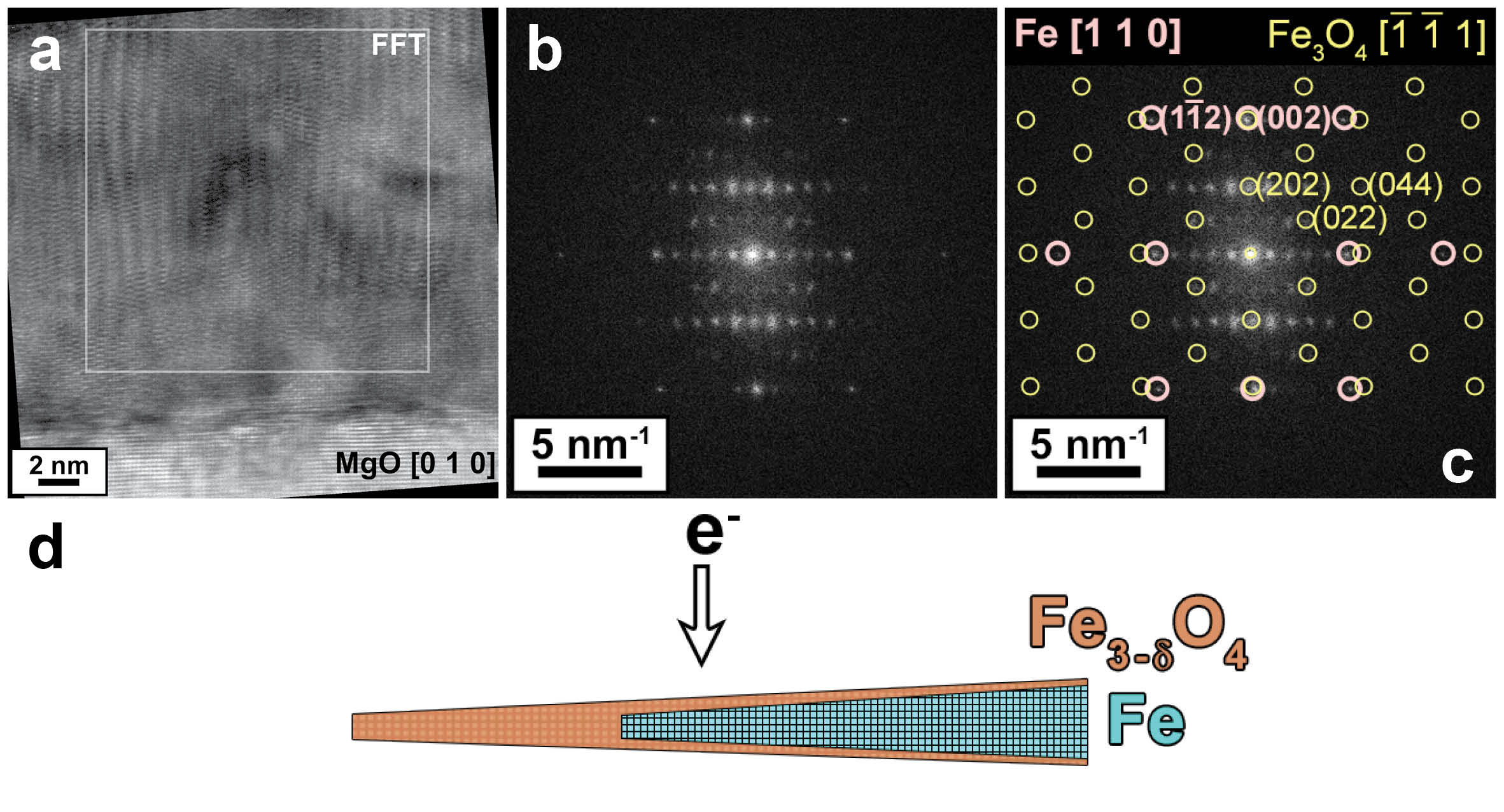} 
	\caption[HRTEM of the iron cross-section]{(Color online) High resolution micrograph of the iron thin film in cross section.  In (a), the HRTEM image is shown along with a box for the region from which an FFT is taken.  The beam is oriented parallel to MgO~[0~1~0] and Fe~[1~1~0].  In (b), the FFT is shown as computed.  In (c), the FFT is indexed using the first order spatial frequencies for Fe and Fe$_3$O$_4$.  The unindexed spatial frequencies can be accounted for through convolution of the two materials, as discussed in the supplementary information.  This information allows us to construct a structural model for the sample in the TEM, which is presented in (d).}
	\label{fig:HRTEM}
\end{figure}

To further refine the structural analysis, High Resolution TEM (HRTEM) experiments were performed on a neighboring region of the film, as displayed in figure \ref{fig:HRTEM}.  The lower convergence angle offered by this technique allows for sharper spots in the Fourier transform of HRTEM images than achievable in the CBED pattern. In figure \ref{fig:HRTEM}a, the MgO substrate appears to be monocrystalline but the iron thin film is not.  The discrete Fast Fourier Transform (FFT) of this image is shown in figure \ref{fig:HRTEM}b, revealing that both high and low frequencies are present.  The lowest order observed spatial frequency $\Delta \mathbf{g}$ is 0.98~nm$^{-1}$, and this is visible in figure \ref{fig:HRTEM}a as a low frequency ``beat'' oriented perpendicular to the substrate.  This $\Delta \mathbf{g}$ vector can be attributed to an interference phenomenon known as Moir{\'e} contrast. The analysis of this phenomena in this manuscript follows the approach described by Amidror \cite{amidror_theory_2009}.  Critically, this approach demonstrates that the best explanation for the oxide structure is the mixed valence Fe$_{3-\delta}$O$_4$ where $\delta$ is close to 0.  If $\delta$ were close to 0.33 yielding a structure more similar to monovalent $\gamma$-Fe$_2$O$_3$ or maghemite, then one would expect additional rows of interference reflections due to the reduction of symmetry arising from the presence of cation vacancies needed to maintain charge neutrality \cite{cornell_iron_2003}.  Other iron oxides can be ruled out due to lacking the appropriate lattice plane spacings and symmetry operations to fit with the experimental observations.  A more detailed analysis including simulations is provided in the SUPPLEMENTARY INFORMATION.  We cannot exclude the possibility that minute amounts of other oxide structures or magnetite with impurities or vacancies may coexist with the Fe$_{3-\delta}$O$_4$ in our sample.  However, these must be present in minute amounts so that they remain undetectable in the HRTEM images and diffraction patterns.  As only large volumes of textured crystals can strongly modify the EMCD signal, we would thus expect the influence of any such phases to be negligible.  

Thus, based on the EELS, CBED, and HRTEM data, it can be concluded that the lamella is best described as a free-standing trilayer of Fe cladded between two thin Fe$_{3-\delta}$O$_4$ layers where $\delta$ is close to 0.  These layers appear to grow with a well-defined texture where Fe~[1~1~0] $\parallel$ Fe$_{3-\delta}$O$_4$ $[\overline{1}~\overline{1}~1]$.  It is worth noting that it has been previously shown that Fe$_{3-\delta}$O$_4$ in this orientation is the most likely oxide to form on an Fe~(1~1~0) surface under similar growth conditions \cite{kim_fe<sub>3</sub>o<sub>4</sub>111/fe110_2000,kim_oxidation_2000}.  A schematic model of this composite structure is presented in figure \ref{fig:HRTEM}d.  

\subsection*{EMCD measurements}

\begin{figure}
	\includegraphics[width=\textwidth]{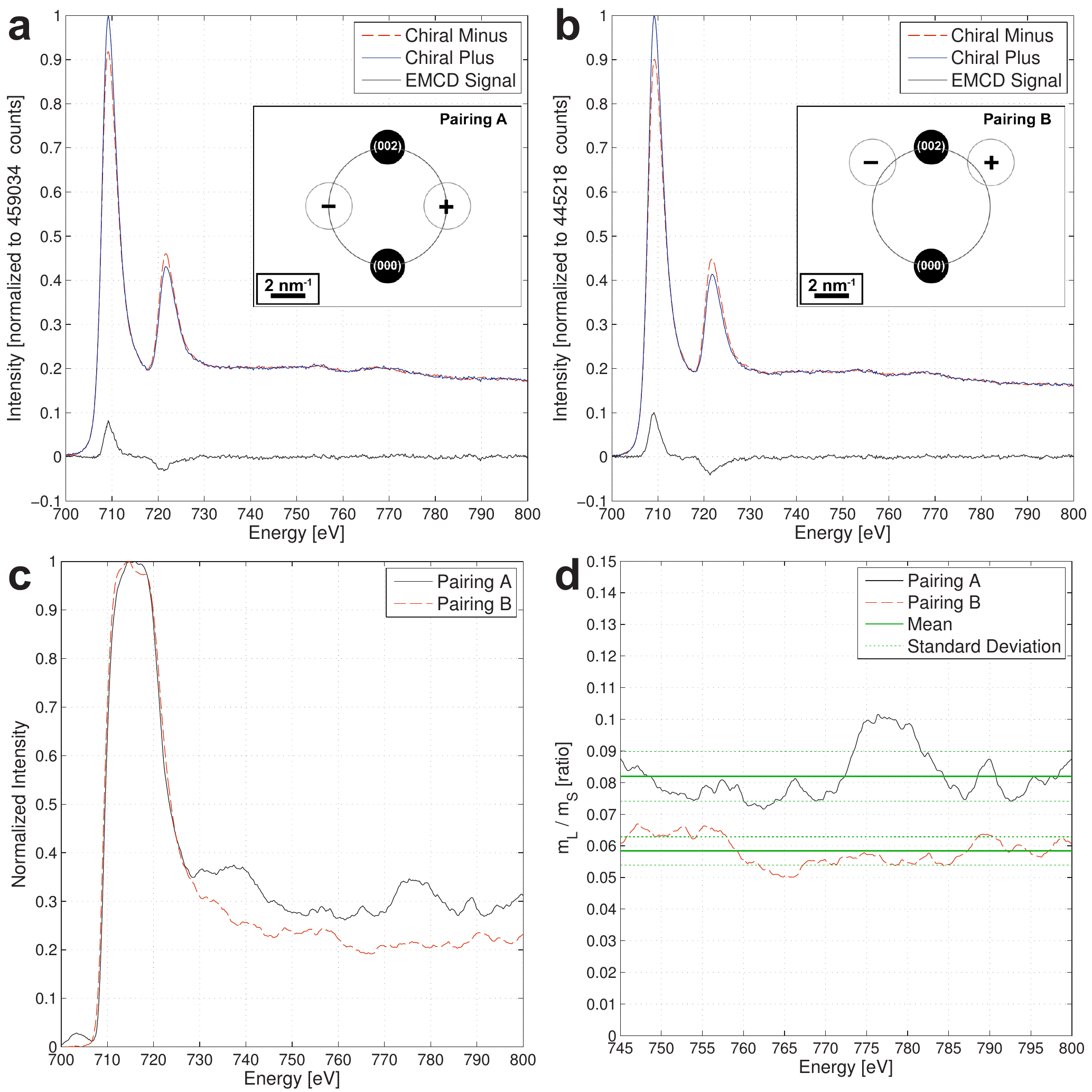} 
	\caption[EMCD measurements in Q-Space]{Background subtracted and post-edge normalized EELS spectra for both chiral locations in aperture pairings A (a) and B (b). Both spectra are normalized to the max value of the Chiral Plus spectra and the normalization factor is provided.  The difference spectrum (EMCD signal) is also presented.  Inset is a schematic depicting the positions of the different aperture pairings in the reciprocal plane. The individual aperture positions are marked + and - to denote Chiral Plus and Chiral Minus, respectively.  (c) Integral of the EMCD signal for both aperture positions. (d) Variation of $m_L / m_S$ as a function of integration range.}
	\label{fig:EMCD}
\end{figure}

The EMCD experiments were performed in a two-beam condition with $\mathbf{g}$ = Fe~(0~0~2), as described in the methods section.  A schematic of the diffraction pattern under these conditions is presented as an inset in figure \ref{fig:EMCD}a,b.  Two regions of q-space were sampled, and these regions are denoted as aperture pairings A and B throughout the text. Aperture pairing A was located on the Thales circle (figure \ref{fig:EMCD}a inset), while pairing B was located closer to the Fe~(0~0~2) reflection (figure \ref{fig:EMCD}b inset).   Each aperture pairing consisted of two individual aperture positions mirrored about the systematic row, thus collecting spectra corresponding to scattering geometries with opposite chirality.  These positions are denoted Chiral Plus and Chiral Minus and are defined for each aperture pairing in the insets of figure \ref{fig:EMCD}a,b.   

Figure \ref{fig:EMCD}a depicts the two chiral EELS spectra acquired using the scattering geometry of aperture pairing A, while figure \ref{fig:EMCD}b shows the corresponding spectra collecting from aperture pairing B.  All spectra in figures \ref{fig:EMCD}a and b are a summation of all of the individual spectra in their respective spectral image and are plotted on a y-axis that has been normalized to the maximum value of the Chiral Plus spectrum in each pairing to facilitate comparison.  In these figures, the pre-edge background for the Chiral Plus and Chiral Minus spectra has been subtracted and the signals have been shifted and aligned with respect to each other as described in the methods section.  They otherwise represent the raw data.  The difference between the two EELS spectra of opposite chirality for each aperture pairing is denominated as the dichroic signal (or ``EMCD signal'') and is shown along with the spectra on the same normalized scale.  The integral of both EMCD signals is presented in figure \ref{fig:EMCD}c.  A clear difference can be seen between the two aperture pairings, indicating that the ratio of the area under the Fe $L_2$ and Fe $L_3$ edges changes between aperture pairings.  When sum rules are applied to the resulting EMCD signals, the value for $m_L / m_S$ can be extracted as described in the methods section and the results for the two aperture pairings are shown in figure \ref{fig:EMCD}d.  The values stabilize after approximately 745~eV and the remaining variations can most likely be attributed to the residual noise.  The average values of $m_L / m_S$ are $0.082 \pm 0.008$ for pairing A and $0.058 \pm 0.004$ for pairing B.  

\begin{figure}
	\includegraphics[width=\textwidth]{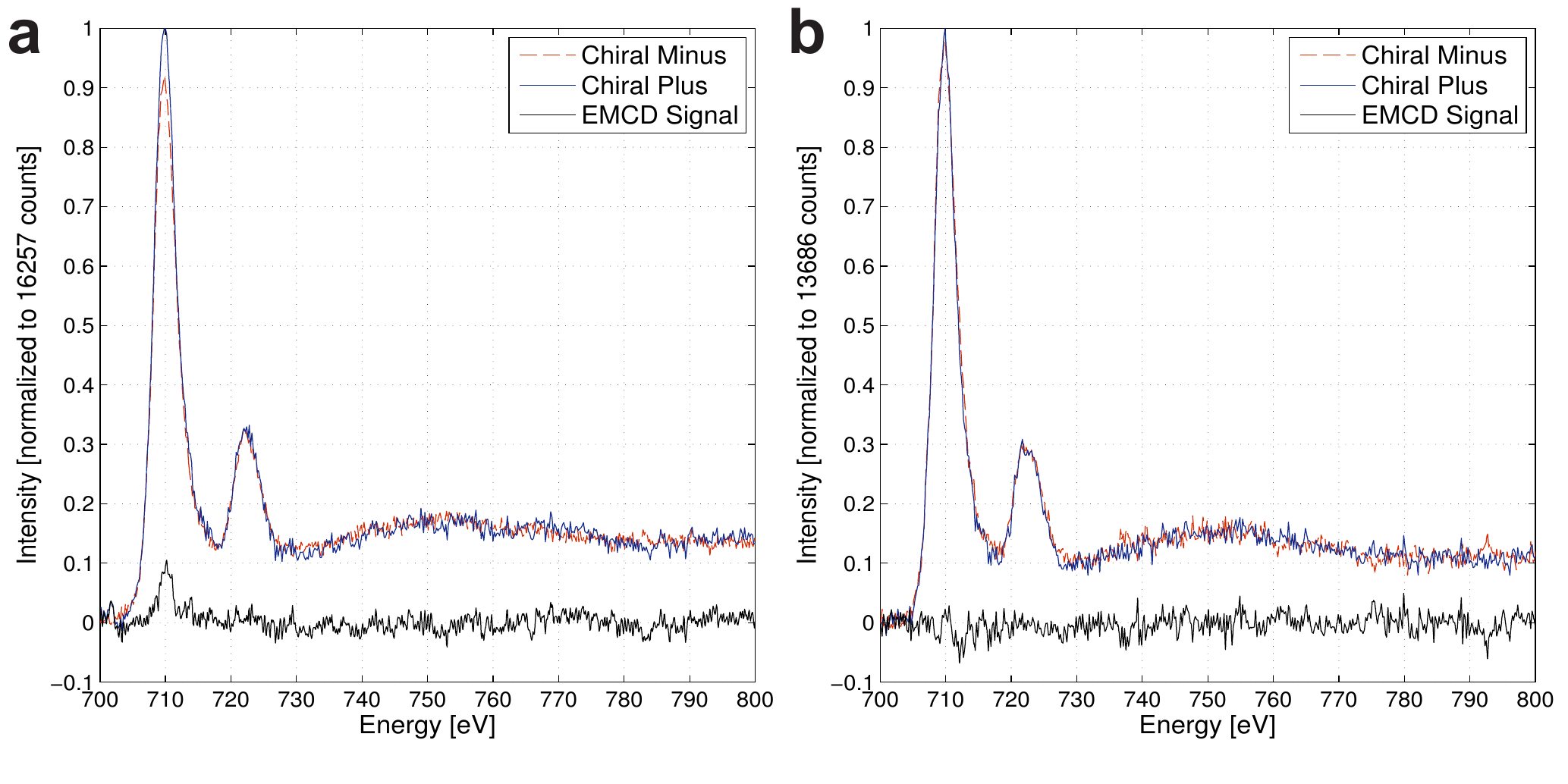} 
	\caption[EMCD from the surface region]{Background subtracted and post-edge normalized EELS spectra for the summation over the ``surface'' region, as denoted in figure 1.  Aperture pairings A and B are presented in (a) and (b), respectively.  The EMCD signals for each chiral pair are also presented.}
	\label{fig:SurfaceEMCD}
\end{figure}

This variation of $m_L / m_S$ in q-space strongly suggests that there are two sources contributing to the measured EMCD signal with dissimilar spin and orbital magnetic components that scatter differently.  The presence of two chemically and structurally distinct layers described in the previous section offers a plausible explanation for this effect.  To more closely correlate this effect to the different layers, we sum the spectra over the ``surface'' region denoted in figure \ref{fig:Survey}.  The results of this summation for aperture pairings A and B are presented in figures \ref{fig:SurfaceEMCD}a and \ref{fig:SurfaceEMCD}b, respectively.  A striking variation is visible.  For aperture pairing A, the asymmetry on the Fe $L_3$ edge is approximately 10\% while on the Fe $L_2$ edge this asymmetry is below the noise level.  This will lead to an increased value of $m_L / m_S$.  This asymmetry nearly vanishes in aperture pairing B, which is additional evidence that the EMCD signal in aperture pairing B is dominated by the metallic iron signal.  

\subsection*{EMCD simulations\label{sec:emcd}}

\begin{figure}
 \includegraphics[width=\textwidth]{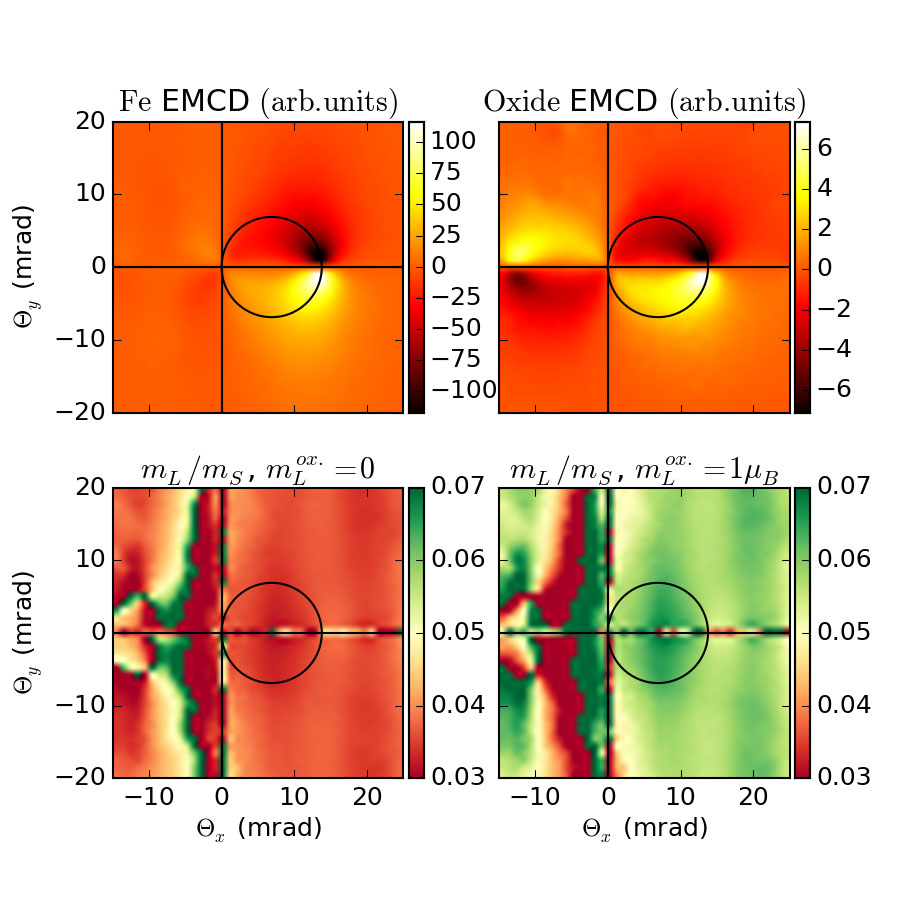}
 \caption[EMCD reciprocal map simulations]{Top row: calculated maps of the EMCD signal at the $L_3$ edge of iron originating from bcc Fe (a) and two 12.5~nm thick oxide layers (b). Bottom row: maps showing the variation of the resulting $m_L/m_S$ ratio, assuming zero orbital angular momentum in oxide (c) or $m_L=1\mu_B$ (d).  The Thales circle is overlaid on all maps.}
 \label{fig:calc}
\end{figure}

To explore potential origins for the q-space dependence of $m_L / m_S$ in this composite system, we simulated its inelastic electron scattering behavior. Figures~\ref{fig:calc}a and b show calculated maps of the magnetic signal originating from $L_3$ edge of iron, separately for the iron layer (fig.~\ref{fig:calc}a) and for the oxide layers (fig.~\ref{fig:calc}b). This signal corresponds to the difference spectrum between chiral minus and chiral plus positions of the $L_3$ edge, as described elsewhere \cite{rusz_local_2011}. It shows that the magnetic signal varies with scattering angles differently for iron than for the oxide layers. Note also the dissimilar order of magnitude of the signals resulting from the different thicknesses and densities of iron atoms between the two layers. 

The $m_L/m_S$ ratio for the experiment can be determined by a linear combination of the results presented in figures \ref{fig:calc}a and b. When sum rules are applied to the resultant composite map, it becomes possible to calculate the variation of $m_L/m_S$ in reciprocal space, and two such maps are presented in figures~\ref{fig:calc}c and d.  Surprisingly, despite the much lower strength of the EMCD signal originating from magnetite, these maps depend quite strongly on the assumed orbital angular momentum of its iron atoms.  Figure \ref{fig:calc}c shows the variation of $m_L/m_S$ ratio under the assumption of zero angular momentum on Fe atoms in the oxide. In this case, the $m_L/m_S$ ratio varies very weakly within the area where measurements were performed, retaining the expected value of $\approx 0.043$. Stronger variations are observed only along the lines where the EMCD signal originating from bcc Fe is negligible, thus even tiny deviations lead to large changes. On the other hand, assuming that the iron atoms in the oxide layer have an enhanced orbital moment (for example $m_L=1\mu_B$), the picture changes substantially (figure~\ref{fig:calc}d). In the area close to the Thales circle (best represented by aperture position A) the value is significantly enhanced, while near the region best represented by aperture position B, the value remains close to the expected $0.043$. The range of this variation scales with the size of orbital angular momentum on oxide iron atoms, with smaller values giving rise to less variation. Therefore, based on our simulations, the large variation of $m_L/m_S$ observed in experiment suggests the presence of large, unquenched orbital angular momenta on the iron atoms in the oxide layer.

\section*{Discussion}
\label{Discussion}

The experimental design presented here provides two ways to study magnetic heterostructures that cannot be performed by any other method.  First, by exploiting the angular dependency of the EMCD signal through the two different aperture pairings, it is possible to experimentally explore the different magnetic scattering contributions for both the metallic iron as well as its oxide surface layer. Second, by scanning the probe over a well-defined area from which individual datasets containing both spatial and spectral information are collected, it becomes possible to spatially segregate the EMCD signal with a spatial resolution of approximately 1.5~nm.

By combining both of these methods, we reach the conclusion that the oxide layer is responsible for the measured changes of $m_L/m_S$.  The simplest explanation could be that this is an artifact in the analysis due to the position and width of the interval used for post-edge normalization of the spectra. Particularly, based on discussion of the influence of magnetic EXAFS in XMCD measurements of magnetite \cite{goering_vanishing_2006}, it was suggested that the normalization window should start above 760~eV. We do observe that if we shrink the normalization interval to a much smaller width of 10eV, the calculated value of $m_L/m_S$ becomes more sensitive to the position of normalization interval. However, note that the post-edge slope is very close to zero both aperture pairings over a wide energy range (see figure \ref{fig:EMCD}c). As a result, as long as the normalization windows for both pairings are kept the same and their width is large enough to mitigate the influence of noise, pairing A will always have a larger value of $m_L/m_S$ than pairing B.  Hence a post-edge normalization artifact can be excluded as the reason for the enhanced $m_L/m_S$ observed in aperture pairing A.

Thus we are led explore the possibility that the orbital component to the net magnetization in the iron oxide layers is enhanced, as suggested by the simulations.  Such results have some precedent in the literature.  In the case of  $\gamma$--Fe$_2$O$_3$, Skoropata et al.~have recently reported a strong increase in the orbital moment for iron in the outer shell of core-shell nanoparticles doped with cobalt \cite{skoropata_magnetism_2014}.  For stoichiometric Fe$_3$O$_4$, Huang et al.~reported orbital moments of $0.67 \pm 0.07~\mu_B$ at all temperatures measured using XMCD \cite{huang_spin_2004}, while Li et al.~inferred a large orbital magnetic moment of $0.51 \pm 0.05~\mu_B$ at 10~K based on direct observation of the spin moment via Compton scattering and a comparison with literature magnetization data \cite{li_spin_2007}.  However, other studies suggest that the orbital moment is quenched within the stoichiometric Fe$_3$O$_4$, resulting in a nearly vanishing net value \cite{duffy_spin_2010, goering_vanishing_2006, goering_absorption_2007}, and these conflicting reports have lead to some controversy \cite{goering_comment_2006, huang_reply_2006}. 

Although our calculations provide strong support for the interpretation suggesting an enhanced orbital angular moment in oxide, we would like to present an alternative argument, which is independent of the simulations. Let's assume that the orbital moments are actually compensated on both iron sublattices and in both iron valencies. The dynamical diffraction effects \cite{warot-fonrose_magnetic_2010,song_effect_2014} will mix the contributions of the three sublattices in non-trivial way. A full disentangling of the individual contributions, as performed by Wang et. al. \cite{wang_quantitative_2013}, is beyond the scope and intentions of this manuscript. However, in general, a linear combination of large spin moment contributions will yield a result larger than a corresponding linear combination of the small orbital moment contributions, especially considering that there is a variation of the thickness of the oxide within the studied region. In such case, the observed $m_L/m_S$ ratio would necessarily remain small. But this is in contradiction with our measurements, implying enhancement of the orbital moment in the oxide layer. On the other hand, it is important to point out that the exact value of $m_L/m_S$ ratio for the oxide-dominated pairing A, or extracted from the oxide-rich area, is difficult to interpret without disentangling the individual site contributions \cite{warot-fonrose_magnetic_2010,wang_quantitative_2013} 

The reason for the variation of orbital magnetic moments in the literature is not always evident, although a number of theories exist.   Kallmayer et al. used a monolayer sensitive XMCD technique to probe the interfaces of epitaxially-grown magnetite thin films on MgO and Al$_2$O$_3$ substrates \cite{kallmayer_magnetic_2008}.  A modest enhancement of the orbital moment was observed at the interface between Fe$_3$O$_4$ and Al$_2$O$_3$.  They interpret this as a consequence of the reduced crystal symmetry of the magnetite at this interface arising from the incorporation of misfit dislocations due to the large lattice mismatch.  Another explanation comes from an investigation into the potential for magnetite to harbor large, hidden orbital moments by E. Goering \cite{goering_large_2011}. Goering calculates an average orbital moment of 1.0 $\mu_{\rm B}$ per Fe atom within magnetite, but that it is nearly completely quenched.  He concludes that ``slight modifications of the stoichiometry and crystallographic structure [of Magnetite] also give nonvanishing orbital moments.''  Since the XMCD technique averages over very large areas of crystalline material, however, it is difficult to isolate these effects.  

Thus it appears that the observed enhancement of the orbital moment may be partially understood by examining the literature.  However, it is critical to emphasize that, due to the unique nature of the EMCD technique presented here, it needs to be considered independently.  For example, to the best of our knowledge, all of the quantitative magnetic information on stoichiometric Fe$_3$O$_4$ to date comes from bulk systems or thin films where the signals are averaged over regions several hundreds of microns in diameter or more.  The results presented here, on the other hand, are a summation of individual spectra acquired from volumes of material illuminated with an electron probe having a diameter of less than 1.5~nm.  Thus nanoscale effects may play a greater role than for measurement techniques that probe much larger volumes of material.  As an example of how this may manifest itself, we note that it is quite plausible that a series of correlated defects at the interface between Fe and magnetite could lead to a reduction in crystalline symmetry that may account for at least some enhancement of the orbital moment, and that the EMCD technique presented here would be exceptionally sensitive to this.  Moreover, although not explicitly investigated in this experiment, the preparation of the lamella with the FIB results in the presence of Gallium impurities on the exposed iron surface.  This may influence the growth and magnetic behavior of the cladding oxide layers.  For example, Gallium ions have been shown to be soluble in magnetite and take on a 3+ oxidation state.  They substitute for Fe$^{3+}$ at the tetrahedral sites, thereby leading to an enhancement of the net magnetic moment \cite{pool_enhanced_2011}.  Although the influence of these impurities on the orbital magnetic moment has not been directly investigated, the suppression of antiferrimagnetic balance of Fe$^{3+}$ ions associated with their introduction may further contribute to the emergence of uncompensated orbital moments.  

In conclusion, we present an EMCD-based method that enables the quantitative analysis of magnetic moments in chemically and structurally distinct overlapping magnetic thin films in the TEM with nanoscale spatial resolution.  Our structural and chemical analysis of the TEM lamella indicate that the system is best described as a free-standing trilayer of Fe$_{3-\delta}$O$_4$~/~Fe~/~Fe$_{3-\delta}$O$_4$ with $\delta$ close to zero.  By angularly and spatially segregating the EMCD signal, we observe significant variations in the measured value of $m_L/m_S$ which can be correlated to the presence of these oxide layers.  Simulations suggest that this can be understood if the orbital component of the net magnetic moment in the oxide is enhanced.  Thus this method is capable of providing significant insight into the nature of nanoscale magnetism in a way not yet possible for any other technique.  

\section*{Methods}
\label{Methods}

\subsection*{Sample Preparation}

A thin film of pure iron was deposited by molecular beam epitaxy onto a single crystal (0~0~1)~MgO substrate.  The sample was prepared for the TEM using the FIB in-situ lift-out method \cite{langford_preparation_2001, langford_situ_2004}.  A protective bar of a Pt-C compound was locally deposited on the region of interest by decomposing a platinum carbon precursor gas in the presence of electrons accelerated to 3~kV.  On top of this, a second bar of approximately 1~$\mu$m thickness was deposited by decomposing the same precursor gas in the presence of gallium ions accelerated to 30~kV.  A lamella roughly 15~$\mu$m in length was milled, extracted, attached to a copper TEM grid using the precursor gas, and thinned to an estimated 100~nm thickness using the 30~kV ion beam.  A final polish using a 5~kV gallium ion beam was completed on both sides with an incidence angle of approximately 3\Deg\ to the lamella face.  

\subsection*{Experimental Equipment}

The TEM and EELS measurements carried out as part of this study were conducted using a Tecnai F30 at Uppsala University as well as a Tecnai F20 at the Argonne National Laboratory Electron Microscopy Center (ANL EMCenter).  Both instruments were equipped with Schottky Field Emission guns and operated at 300~kV and 200~kV, respectively.  EELS data were acquired on a Tridiem Gatan Image Filter (GIF) (Gatan Inc.).  The field emission gun was operated in such a way to produce a high current at the expense of energy resolution, which was close to 1.3~eV, taken as the Full Width Half Max (FWHM) value of the Zero Loss Peak (ZLP).  All EELS spectra were acquired with the microscope set in Scanning TEM (STEM) mode with a calibrated camera length of 1877~mm.  The minicondenser lens was switched off to yield a lower convergence angle of 1.6~mrad for a fully converged beam, which was diffraction limited by use of the secondary condenser aperture.  We estimate that the spatial resolution of this configuration is 1.2~nm.  The EELS entrance aperture (physical diameter 1.0~mm) was used to set the collection angle at 2.4~mrad.  The aperture was positioned using a script to excite the diffraction shift coils and the exact positions are shown in the insets of figure \ref{fig:EMCD}a,b.

In STEM mode, a High Angle Annular Dark Field (HAADF) detector was used to produce a survey image using electrons scattered by angles larger than approximately 30~mrad.  This resulted primarily in a mass-thickness contrast mechanism, but also included some diffraction contrast.  The region from which all spectrum image data cubes were acquired is shown in green in figure \ref{fig:Survey}.  A drift correction routine carried out at regular intervals ensured that the probe position within this region could be linked to the survey image and, subsequently, related between all of the individual data cubes.  The probe was scanned perpendicular to the Fe/MgO interface with a pixel size of approximately 1.5 $\times$ 1.5~nm.  

The EMCD data were acquired by first shifting the aperture to the ``Chiral Plus'' location of the pairing A shown in the inset of figure \ref{fig:EMCD}a and recording the energy range 480 - 890~eV using an acquisition time of 5~s per pixel and an energy dispersion of 0.2~eV per channel.  The aperture position was subsequently shifted to the ``Chiral Minus'' location and the same region was scanned again.  This was repeated for both aperture positions in pairing B (see figure \ref{fig:EMCD}b, inset).  For the core-loss EELS data, the diffraction pattern was shifted on-axis and the acquisition time was reduced to 2~s per pixel.  The low-loss region was acquired with the diffraction pattern on-axis using a dispersion of 0.05~eV and an acquisition time of $1\times 10^{-2}$~s.  Critically, the survey image was not reacquired between aperture shifts, allowing for the same region to be scanned multiple times.  Following the acquisition of each individual spectral image, correlated noise was accounted for by taking the average of $3\sqrt{N}$ dark current measurements - where $N$ is the total number of acquisitions - and subsequently subtracted from the gain normalized spectra \cite{hou_reduce_2009}.  

\subsection*{Data Treatment}

Energy drift in the individual spectra within the spectral images was corrected for by using a cross-correlation algorithm to align the spectra to the Fe~$L_3$ ionization edge within the regions of the film having the largest metallic iron content.  The drift correction for the remaining spectra was determined by interpolating a spline fit between the non-corrected regions.  For the EMCD spectra, the energy drift correction resulted in a gain averaging over approximately 15 channels, further improving the signal to noise ratio \cite{bosman_optimizing_2008}.  Following this step, the effects of plural scattering were removed by deconvolving all of the core-loss spectral images with the on-axis low-loss spectral image \cite{rusz_influence_2011}.  The individual spectra were then summed into two datasets for each aperture position: the whole image and the ``surface'' region (see figure \ref{fig:Survey}).  Following the summation, the pre-edge background was removed using a power-law background fit to the region between 670 -- 700~eV \cite{egerton_electron_2011}.  

The on-axis core-loss spectral image was used to quantify the relative amount of iron $r_{\rm{Fe}}$ and oxygen $r_{\rm{ox}}$ at any given pixel.  This was accomplished by fitting the deconvolved data to the differential scattering cross-sections calculated using the Hartree-Slater model as implemented in Digital Micrograph.  The effective electron mean free path was then computed for each pixel by utilizing the Fe:O ratio $R = r_{\rm{Fe}} / r_{\rm{ox}}$ at that position as proposed by Malis \textit{et al}.~\cite{malis_eels_1988} and implemented by Egerton in Matlab \cite{egerton_electron_2011}.  This was used to calculate the absolute thickness $t_{\rm{abs}}$ of the region at each pixel position by using the low-loss spectral image to first extract the relative thickness values.  An equation governing the thickness of the oxide layer $t_{\textup{ox}}$ as a function of $R$ and $t_{\rm{abs}}$ was derived under the assumption of it being similar in structure to Fe$_3$O$_4$.  The justification for this assumption is provided in the supplementary information. This yields equation \ref{eq:OxideThickness}
\begin{equation}
t_{\textup{ox}} = \frac{t_{\rm{abs}} \rho _{\rm{Fe}}}{\frac{4R}{7}\rho _{\rm{ox}}-\frac{3}{7}\rho_{\rm{ox}}+\rho_{\rm{Fe}}}
\label{eq:OxideThickness}
\end{equation}
where  $\rho_{\textup{ox}}$ and $\rho_{\textup{Fe}}$ are the densities of the oxide and the metallic Fe layers, respectively.  $t_{\rm{Fe}}$ was calculated as $t_{\rm{abs}} - t_{\rm{ox}}$.  The primary sources of systematic error for this calculation include the choice of cross-section model and the quantification routine, uncertainty in the local densities of the oxide and metal, the potential presence of an amorphous carbon coating layer, and unknown stoichiometric deviations from pure Fe$_3$O$_4$.  Since many of these error sources are difficult to quantify, we are not able to provide systematic error bars for this calculation.  Despite that, we note that the statistical error appears to be quite low due to the high signal to noise ratio.  Consequently, we feel that these data provide a constructive qualitative assessment of the thicknesses of the individual metal and oxide layers.

The EMCD signal was computed by first interpolating the background-removed data to a dispersion of 0.01~eV and then aligning the two spectra of opposite chirality from each aperture pairing along the energy dispersion axis with a cross-correlation algorithm.  The post-edge background was subsequently normalized to a window between 745 -- 800~eV and the difference between the spectra was computed.  With this, the conditions for sum rules are satisfied and $m_L / m_S$ can be calculated using equation \ref{eq:SumRules}.

\begin{equation}
\frac{m_L}{m_S} = \frac{2}{3}\frac{\int_{\rm{L3}}^{~}\Delta I(E)dE + \int_{\rm{L2}}^{~}\Delta I(E)dE}{\int_{\rm{L3}}^{~}\Delta I(E)dE -2 \int_{\rm{L2}}^{~}I(E)dE}
\label{eq:SumRules}
\end{equation}

The $m_L / m_S$ values in this paper were determined from the difference in the background-removed post-edge normalized data.  The integration range for Fe $L_3$ was fixed at 700 -- 715~eV. The lower bound of the integration range for Fe $L_2$ was set at 700~eV and the upper bound was varied from 745~eV up to 800~eV.  This variation is shown in figure \ref{fig:EMCD}f.  The reported values for $m_L / m_S$ represent the average value of this variation between 745 -- 800~eV while the error bars represent one standard deviation.

We note that this way of processing data, particularly the post-edge normalization step before taking the difference, significantly suppresses the effects of asymmetry of the two beam case \cite{rusz_asymmetry_2010,lidbaum_reciprocal_2010,song_effect_2014}. This follows from the cubic symmetry of both the magnetite and iron layer, which allows to write each ELNES spectrum as a linear combination of nonmagnetic spectrum $N(E)$ and magnetic EMCD part $M_{z}(E)$:

\begin{equation}
\frac{\partial^2 \sigma\left ( E,\Omega \right )}{\partial E \partial \Omega}\propto A(\Omega)N(E) + B(\Omega)M_{z}(E)
\end{equation}

The post-edge normalization removes the differences in geometry-dependent coefficients $A(\Omega)$, because EMCD is negligible in the post-edge region. Thus the difference of post-edge normalized spectra faithfully represents the EMCD spectrum.

\subsection*{Simulations\label{sec:simul}}

Simulations of the inelastic electron scattering were performed using the Bloch-waves method \cite{saldin_theory_1987,rossouw_implications_1984,rusz_first-principles_2007} utilizing the \textsc{mats} algorithm \cite{rusz_new_2013}. The orientations and thicknesses of the layers followed the observed ones. The transition matrix elements--mixed dynamical form factors \cite{kohl_theory_1985} were calculated using the operator maps technique \cite{rusz_local_2011}, which allows for easy scaling of the magnitude of spin and orbital angular momenta \cite{rusz_sum_2007} at the cost of being a dipole approximation. For bcc Fe, we used the values $m_S=1.98\mu_B$ and $m_L=0.086\mu_B$ as reported by Chen et al. \cite{chen_experimental_1995}. For magnetite layers, we assumed spin moments of $m_S=5.0\mu_B$ and $m_S=4.0\mu_B$, with a net magnetization of $m_S=4.0\mu_B$.  The value for $m_L$ was set to either zero (figure \ref{fig:calc}c) or $1\mu_B$ (figure \ref{fig:calc}d). The moments on the two iron sublattices of magnetite were oriented ferrimagnetically, assuming that its net magnetic moment is parallel with that of the iron layer. Resulting maps were scaled to take into account the thickness of each layer and amount of iron atoms per unit volume in iron (84.6~at./nm$^3$) and magnetite (13.5 and 26.99~at./nm$^3$), respectively.  Note the different magnitude of magnetic signals from the $L_3$ edges in the iron (figure \ref{fig:calc}a) and oxide (figure \ref{fig:calc}b) layers, respectively.  We have also performed test calculations comparing plane wave illumination and convergent beam illumination with a 1.6~mrad convergence angle beam in the (0~0~1) zone axis condition for bcc--Fe. This orientation was chosen so that the dynamical diffraction effects are maximized, thus providing an upper limit on the differences that could be expected between the two methods of calculation. However, thanks to Lorentzian broadening of the energy-filtered diffraction patterns due to energy loss process \cite{egerton_electron_2011}, the resulting diffraction patterns and EMCD were very similar.  Thus for the multilayer sample, we proceeded with simpler plane-wave simulations.

\section*{Acknowledgements}
The authors acknowledge the STINT research grant (1G2009-2017).  In addition, part of this study was accomplished at the Electron Microscopy Center at Argonne National Laboratory, a U.S. Department of Energy Office of Science Laboratory operated under Contract No. DE-AC02-06CH11357 by UChicago Argonne, LLC. 
J.~R. acknowledges the Swedish Research Council, G\"{o}ran Gustafsson's Foundation and Swedish National Infrastructure for Computing (NSC center).  Special thanks are also due to Paul Thomas at Gatan Inc.~for providing the authors with an automated procedure to execute the high quality dark reference correction.

\section*{Author contributions}
T.~T. designed and executed the experiment, analyzed and interpreted the results, was the main author of the manuscript, and developed the code for quantifying the EMCD spectra.  K.~L. supervised the experiment and contributed to the writing of the manuscript as well as the analysis of the data.  J.~R. developed the simulations, interpreted the results, and contributed to the writing of the manuscript.  S.~R. contributed to the interpretation of the Moir{\'e} contrast. B.~H. provided the sample and assisted in the conceptualization of the experiment.  Y.~I. and N.~Z. assisted with the experimental design and execution.  All authors discussed the results and commented on the manuscript.

\section*{Additional information}
\textbf{Competing financial interests:} The authors declare no competing financial interests.


\begin{thebibliography}{10}

\bibitem{gambardella_ferromagnetism_2002}
Gambardella, P. \emph{et~al.}
\newblock Ferromagnetism in one-dimensional monatomic metal chains.
\newblock \emph{Nature} \textbf{416}, 301--304 (2002).

\bibitem{katsnelson_half-metallic_2008}
Katsnelson, M.~I., Irkhin, V.~Y., Chioncel, L., Lichtenstein, A.~I., and
  de~Groot, R.~A.
\newblock Half-metallic ferromagnets: From band structure to many-body effects.
\newblock \emph{Rev. Mod. Phys.} \textbf{80}, 315--378 (2008).

\bibitem{bibes_oxide_2007}
Bibes, M. and Barthelemy, A.
\newblock Oxide Spintronics.
\newblock \emph{{IEEE} Trans. Electron Dev.} \textbf{54}, 1003--1023 (2007).

\bibitem{wada_efficient_2010}
Wada, E., Watanabe, K., Shirahata, Y., Itoh, M., Yamaguchi, M., and Taniyama,
  T.
\newblock Efficient spin injection into {GaAs} quantum well across
  Fe$_{\textrm{3}}$O$_{\textrm{4}}$ spin filter.
\newblock \emph{Appl. Phys. Lett.} \textbf{96}, 102510 (2010).

\bibitem{monti_magnetism_2012}
Monti, M. \emph{et~al.}
\newblock Magnetism in nanometer-thick magnetite.
\newblock \emph{Phys. Rev. B} \textbf{85} (2012).

\bibitem{duffy_spin_2010}
Duffy, J.~A. \emph{et~al.}
\newblock Spin and orbital moments in Fe$_{\textrm{3}}$O$_{\textrm{4}}$.
\newblock \emph{Phys. Rev. B} \textbf{81} (2010).

\bibitem{goering_vanishing_2006}
Goering, E., Gold, S., Lafkioti, M., and Schütz, G.
\newblock Vanishing Fe 3d orbital moments in single-crystalline magnetite.
\newblock \emph{Europhys. Lett.} \textbf{73}, 97--103 (2006).

\bibitem{goering_absorption_2007}
Goering, E., Lafkioti, M., Gold, S., and Schuetz, G.
\newblock Absorption spectroscopy and {XMCD} at the Verwey transition of
  Fe$_{\textrm{3}}$O$_{\textrm{4}}$.
\newblock \emph{J. Magn. Magn. Mater.} \textbf{310}, e249--e251 (2007).

\bibitem{pellegrin_characterization_1999}
Pellegrin, E. \emph{et~al.}
\newblock Characterization of Nanocrystalline
  $\gamma$-Fe$_{\textrm{2}}$O$_{\textrm{3}}$ with Synchrotron Radiation
  Techniques.
\newblock \emph{Phys. Status Solidi B} \textbf{215}, 797--801 (1999).

\bibitem{huang_spin_2004}
Huang, D.~J. \emph{et~al.}
\newblock Spin and Orbital Magnetic Moments of
  Fe$_{\textrm{3}}$O$_{\textrm{4}}$.
\newblock \emph{Phys. Rev. Lett.} \textbf{93} (2004).

\bibitem{li_spin_2007}
Li, Y., Montano, P., Barbiellini, B., Mijnarends, P., Kaprzyk, S., and Bansil,
  A.
\newblock Spin moment over 10-300K and delocalization of magnetic electrons
  above the Verwey transition in magnetite.
\newblock \emph{J. Phys. Chem. Solids} \textbf{68}, 1556--1560 (2007).

\bibitem{goering_large_2011}
Goering, E.
\newblock Large hidden orbital moments in magnetite.
\newblock \emph{Phys. Status Solidi B} \textbf{248}, 2345--2351 (2011).

\bibitem{skoropata_magnetism_2014}
Skoropata, E., Desautels, R.~D., Chi, C.-C., Ouyang, H., Freeland, J.~W., and
  van Lierop, J.
\newblock Magnetism of iron oxide based core-shell nanoparticles from interface
  mixing with enhanced spin-orbit coupling.
\newblock \emph{Phys. Rev. B} \textbf{89} (2014).

\bibitem{hebert_proposal_2003}
Hébert, C. and Schattschneider, P.
\newblock A proposal for dichroic experiments in the electron microscope.
\newblock \emph{Ultramicroscopy} \textbf{96}, 463--468 (2003).

\bibitem{schattschneider_detection_2006}
Schattschneider, P. \emph{et~al.}
\newblock Detection of magnetic circular dichroism using a transmission
  electron microscope.
\newblock \emph{Nature} \textbf{441}, 486--488 (2006).

\bibitem{schattschneider_detection_2008}
Schattschneider, P. \emph{et~al.}
\newblock Detection of magnetic circular dichroism on the two-nanometer scale.
\newblock \emph{Phys. Rev. B} \textbf{78} (2008).

\bibitem{lidbaum_reciprocal_2010}
Lidbaum, H. \emph{et~al.}
\newblock Reciprocal and real space maps for {EMCD} experiments.
\newblock \emph{Ultramicroscopy} \textbf{110}, 1380--1389 (2010).

\bibitem{schattschneider_energy_2008}
Schattschneider, P. \emph{et~al.}
\newblock Energy loss magnetic chiral dichroism: A new technique for the study
  of magnetic properties in the electron microscope (invited).
\newblock \emph{J. Appl. Phys.} \textbf{103}, 07D931 (2008).

\bibitem{che_characterization_2011}
Che, R.~C., Liang, C.~Y., He, X., Liu, H.~H., and Duan, X.~F.
\newblock Characterization of magnetic domain walls using electron magnetic
  chiral dichroism.
\newblock \emph{Sci. Technol. Adv. Mat.} \textbf{12}, 025004 (2011).

\bibitem{stoger-pollach_emcd_2011}
Stöger-Pollach, M., Treiber, C., Resch, G., Keays, D., and Ennen, I.
\newblock {EMCD} real space maps of Magnetospirillum magnetotacticum.
\newblock \emph{Micron} \textbf{42}, 456--460 (2011).

\bibitem{carretero-genevrier_chemical_2012}
Carretero-Genevrier, A. \emph{et~al.}
\newblock Chemical synthesis of oriented ferromagnetic {LaSr}-2 x 4 manganese
  oxide molecular sieve nanowires.
\newblock \emph{Chem. Commun.} \textbf{48}, 6223 (2012).

\bibitem{loukya_electron_2012}
Loukya, B., Zhang, X., Gupta, A., and Datta, R.
\newblock Electron magnetic chiral dichroism in {CrO}$_{\textrm{2}}$ thin films
  using monochromatic probe illumination in a transmission electron microscope.
\newblock \emph{J. Magn. Magn. Mater.} \textbf{324}, 3754--3761 (2012).

\bibitem{salafranca_surfactant_2012}
Salafranca, J. \emph{et~al.}
\newblock Surfactant Organic Molecules Restore Magnetism in Metal-Oxide
  Nanoparticle Surfaces.
\newblock \emph{Nano Lett.} \textbf{12}, 2499--2503 (2012).

\bibitem{warot-fonrose_magnetic_2010}
Warot-Fonrose, B., Gatel, C., Calmels, L., Serin, V., Snoeck, E., and Cherifi,
  S.
\newblock Magnetic properties of {FeCo} alloys measured by energy-loss magnetic
  chiral dichroism.
\newblock \emph{J. Appl. Phys.} \textbf{107}, 09D301 (2010).

\bibitem{calmels_atomic_2011}
Calmels, L. and Rusz, J.
\newblock Atomic site sensitivity of the energy loss magnetic chiral dichroic
  spectra of complex oxides.
\newblock \emph{J. Appl. Phys.} \textbf{109}, 07D328 (2011).

\bibitem{wang_quantitative_2013}
Wang, Z.~Q., Zhong, X.~Y., Yu, R., Cheng, Z.~Y., and Zhu, J.
\newblock Quantitative experimental determination of site-specific magnetic
  structures by transmitted electrons.
\newblock \emph{Nat. Commun.} \textbf{4}, 1395 (2013).

\bibitem{rubino_energy-loss_2008}
Rubino, S. \emph{et~al.}
\newblock Energy-loss magnetic chiral dichroism ({EMCD}): Magnetic chiral
  dichroism in the electron microscope.
\newblock \emph{J. Mater. Res.} \textbf{23}, 2582--2590 (2008).

\bibitem{hebert_magnetic_2008}
Hébert, C., Schattschneider, P., Rubino, S., Novak, P., Rusz, J., and
  Stöger-Pollach, M.
\newblock Magnetic circular dichroism in electron energy loss spectrometry.
\newblock \emph{Ultramicroscopy} \textbf{108}, 277--284 (2008).

\bibitem{rubino_simulation_2010}
Rubino, S., Schattschneider, P., Rusz, J., Verbeeck, J., and Leifer, K.
\newblock Simulation of magnetic circular dichroism in the electron microscope.
\newblock \emph{J. Phys. D: Appl. Phys.} \textbf{43}, 474005 (2010).

\bibitem{rusz_sum_2007}
Rusz, J., Eriksson, O., Novák, P., and Oppeneer, P.~M.
\newblock Sum rules for electron energy loss near edge spectra.
\newblock \emph{Phys. Rev. B} \textbf{76}, 060408 (2007).

\bibitem{calmels_experimental_2007}
Calmels, L. \emph{et~al.}
\newblock Experimental application of sum rules for electron energy loss
  magnetic chiral dichroism.
\newblock \emph{Phys. Rev. B} \textbf{76}, 060409 (2007).

\bibitem{lidbaum_quantitative_2009}
Lidbaum, H. \emph{et~al.}
\newblock Quantitative Magnetic Information from Reciprocal Space Maps in
  Transmission Electron Microscopy.
\newblock \emph{Phys. Rev. Lett.} \textbf{102} (2009).

\bibitem{rusz_local_2011}
Rusz, J., Rubino, S., Eriksson, O., Oppeneer, P., and Leifer, K.
\newblock Local electronic structure information contained in energy-filtered
  diffraction patterns.
\newblock \emph{Phys. Rev. B} \textbf{84} (2011).

\bibitem{rusz_influence_2011}
Rusz, J. \emph{et~al.}
\newblock Influence of plural scattering on the quantitative determination of
  spin and orbital moments in electron magnetic chiral dichroism measurements.
\newblock \emph{Phys. Rev. B} \textbf{83} (2011).

\bibitem{amidror_theory_2009}
Amidror, I.
\newblock \emph{The theory of the Moiré phenomenon}.
\newblock Number 38 in Computational imaging and vision.
\newblock (Kluwer Academic, Dordrecht, 2009).

\bibitem{cornell_iron_2003}
Cornell, R.~M.
\newblock \emph{The iron oxides: structure, properties, reactions, occurrences,
  and uses}.
\newblock (Wiley-{VCH}, Weinheim, 2003).

\bibitem{kim_fe<sub>3</sub>o<sub>4</sub>111/fe110_2000}
Kim, H.-J., Park, J.-H., and Vescovo, E.
\newblock Fe$_{\textrm{3}}$O$_{\textrm{4}}$(111)/Fe(110) magnetic
  bilayer: Electronic and magnetic properties at the surface and interface.
\newblock \emph{Phys. Rev. B} \textbf{61}, 15288--15293 (2000).

\bibitem{kim_oxidation_2000}
Kim, H.-J., Park, J.-H., and Vescovo, E.
\newblock Oxidation of the Fe (110) surface: An
  Fe$_{\textrm{3}}$O$_{\textrm{4}}$ (111) / Fe (110) bilayer.
\newblock \emph{Phys. Rev. B} \textbf{61}, 15284--15287 (2000).

\bibitem{goering_comment_2006}
Goering, E., Lafkioti, M., and Gold, S.
\newblock Comment on “Spin and Orbital Magnetic Moments of
  Fe$_{\textrm{3}}$O$_{\textrm{4}}$”.
\newblock \emph{Phys. Rev. Lett.} \textbf{96} (2006).

\bibitem{huang_reply_2006}
Huang, D.~J., Lin, H.-J., and Chen, C.~T.
\newblock Reply to comment on “Spin and Orbital Magnetic Moments of
  Fe$_{\textrm{3}}$O$_{\textrm{4}}$”.
\newblock \emph{Phys. Rev. Lett.} \textbf{96} (2006).

\bibitem{kallmayer_magnetic_2008}
Kallmayer, M. \emph{et~al.}
\newblock Magnetic moment investigations of epitaxial magnetite thin films.
\newblock \emph{J. Appl. Phys.} \textbf{103}, 07D715 (2008).

\bibitem{pool_enhanced_2011}
Pool, V.~L., Klem, M.~T., Chorney, C.~L., Arenholz, E.~A., and Idzerda, Y.~U.
\newblock Enhanced magnetism of Fe$_{\textrm{3}}$O$_{\textrm{4}}$ nanoparticles
  with Ga doping.
\newblock \emph{J. Appl. Phys.} \textbf{109}, 07B529 (2011).

\bibitem{langford_preparation_2001}
Langford, R.~M. and Petford-Long, A.~K.
\newblock Preparation of transmission electron microscopy cross-section
  specimens using focused ion beam milling.
\newblock \emph{J. Vac. Sci. Technol.} \textbf{19}, 2186--2193 (2001).

\bibitem{langford_situ_2004}
Langford, R.~M. and Clinton, C.
\newblock In situ lift-out using a {FIB}-{SEM} system.
\newblock \emph{Micron} \textbf{35}, 607--611 (2004).

\bibitem{hou_reduce_2009}
Hou, V.-D.
\newblock Reduce Correlated Noise in {EELS} Spectrum with High Quality Dark
  Reference.
\newblock \emph{Microsc. Microanal.} \textbf{15}, 226--227 (2009).

\bibitem{bosman_optimizing_2008}
Bosman, M. and Keast, V.~J.
\newblock Optimizing {EELS} acquisition.
\newblock \emph{Ultramicroscopy} \textbf{108}, 837--846 (2008).

\bibitem{egerton_electron_2011}
Egerton, R.~F.
\newblock \emph{Electron energy-loss spectroscopy in the electron microscope}.
\newblock (Springer, 2011).

\bibitem{malis_eels_1988}
Malis, T., Cheng, S.~C., and Egerton, R.~F.
\newblock {EELS} log-ratio technique for specimen-thickness measurement in the
  {TEM}.
\newblock \emph{J. Elec. Microsc. Tech.} \textbf{8}, 193--200 (1988).

\bibitem{rusz_asymmetry_2010}
Rusz, J., Oppeneer, P., Lidbaum, H., Rubino, S., and Leifer, K.
\newblock Asymmetry of the two-beam geometry in {EMCD} experiments.
\newblock \emph{J. Microsc.} \textbf{237}, 465--468 (2010).

\bibitem{saldin_theory_1987}
Saldin, D.~K.
\newblock The theory of electron energy-loss near-edge structure.
\newblock \emph{Philos. Mag.} \textbf{56}, 515--525 (1987).

\bibitem{rossouw_implications_1984}
Rossouw, C.~J. and Maslen, V.~M.
\newblock Implications of (e, 2e) scattering for inelastic electron diffraction
  in crystals {II}. Application of the theory.
\newblock \emph{Philos. Mag} \textbf{49}, 743--757 (1984).

\bibitem{rusz_first-principles_2007}
Rusz, J., Rubino, S., and Schattschneider, P.
\newblock First-principles theory of chiral dichroism in electron microscopy
  applied to 3d ferromagnets.
\newblock \emph{Phys. Rev. B} \textbf{75}, 214425 (2007).

\bibitem{rusz_new_2013}
Rusz, J., Muto, S., and Tatsumi, K.
\newblock New algorithm for efficient Bloch-waves calculations of
  orientation-sensitive {ELNES}.
\newblock \emph{Ultramicroscopy} \textbf{125}, 81--88 (2013).

\bibitem{kohl_theory_1985}
Kohl, H. and Rose, H.
\newblock Theory of Image Formation by Inelastically Scattered Electrons in the
  Electron Microscope.
\newblock \emph{Advances in Electronics and Electron Physics}, Hawkes, P. (ed.), 173--227 (Academic Press 1985).

\bibitem{chen_experimental_1995}
Chen, C.~T. \emph{et~al.}
\newblock Experimental Confirmation of the X-Ray Magnetic Circular Dichroism
  Sum Rules for Iron and Cobalt.
\newblock \emph{Phys. Rev. Lett.} \textbf{75}, 152--155 (1995).

\end{thebibliography}

\end{document}